\documentclass[twocolumn,showpacs,preprintnumbers,amsmath,amssymb]{revtex4}
\usepackage{dcolumn}
\usepackage{bm}
\usepackage{graphicx}
\usepackage{xcolor}

\usepackage{hyperref}
\hypersetup{colorlinks,
linkcolor=blue,
filecolor=green,
urlcolor=blue,
citecolor=blue}

\def\wc{{\omega_c}}
\def\bE{{\bf E}}
\def\wo{{\omega_0}}
\def\wk{{\omega_k}}
\def\wg{{\omega_g}}
\def\akl{a_{\bk \lambda}}

\def\bd{{\bf d}}
\def\bk{{\bf k}}
\def\ekl{\hat{e}_{\bk \lambda}}
\def\bd{{\bf d}}
\def\bp{{\bf p}}
\def\bR{{\bf R}}

\def\bE{{\bf E}}

\def\bp{{\bf p}}
\def\k0{k_0}

\def\bR{{\bf R}}

\def\akl{a_{\bk \lambda}}

\begin{document}

\title{Control of spontaneous emission of a single quantum emitter through a time-modulated photonic-band-gap environment}

\author{Giuseppe Calaj\`{o}$^{1}$\footnote{giuseppe.calajo@tuwien.ac.at},
Lucia Rizzuto$^{2,3}$\footnote{lucia.rizzuto@unipa.it},
and Roberto Passante$^{2,3}$\footnote{roberto.passante@unipa.it}}
\affiliation{$^1$Vienna Center for Quantum Science and Technology, Atominstitut, TU Wien, 1040 Vienna, Austria}
\address{$^2$Dipartimento di Fisica e Chimica, Universit\`{a} degli Studi di Palermo and CNISM, Via Archirafi 36, I-90123 Palermo, Italy}
\address{$^3$INFN, Laboratori Nazionali del Sud, I-95123 Catania, Italy}

\begin{abstract}
We consider the spontaneous emission of a two-level quantum emitter, such as an atom or a quantum dot, in a modulated time-dependent environment with a photonic band gap. An example of such an environment is a dynamical photonic crystal or any other environment with a bandgap whose properties are modulated in time, in the effective mass approximation. After introducing our model of dynamical photonic crystal, we show that it allows new possibilities to control and tailor the physical features of the emitted radiation, specifically its frequency spectrum. In the weak coupling limit and in an adiabatic case, we obtain the emitted spectrum and we show the appearance of two lateral peaks due to the presence of the modulated environment, separated from the central peak by the modulation frequency.
We show that the two side peaks are not symmetric in height, and that their height ratio can be exploited to investigate the density of states of the environment.
Our results show that a dynamical environment can give further possibilities to modify the spontaneous emission features, such as its spectrum and emission rate, with respect to a static one. Observability of the phenomena we obtain is discussed, as well as relevance for tailoring and engineering radiative processes.

\end{abstract}

\maketitle

\section{Introduction}
\label{sec:Intoduction}

The possibility of controlling and tailoring radiative processes of atoms and quantum emitters \cite{LMS15}, in particular the spontaneous emission process, through the properties of the environment has been investigated since the pioneering works of Purcell \cite{Purcell46}, Kleppner \cite{Kleppner81}, Yablonovitch \cite{Yablonovitch87}, John and Quang \cite{JQ94}  and, more recently, many others (for a review, see \cite{AKP04,LNNB00,LMS15}).
This possibility has also received remarkable experimental verifications \cite{NFA07,Thompson13}.
Photonic-band-gap environments, for example  photonic crystals \cite{LMS15,JVF97,Goban14}, nanophotonic waveguides \cite{NH12,arcari}, coupled cavity arrays \cite{HBP08} and coupled transmission line resonators \cite{Houck} have shown excellent potentialities to control and tailor radiative properties of atoms, molecules and quantum dots.
Photonic crystals are structured materials made by a periodic arrangement of dielectrics with different refractive index \cite{JJWM08}.
Coupled cavity arrays are arrays of cavities where photon hopping can occur between neighboring cavities \cite{HBP08}.
Important examples are inhibition and enhancement of the spontaneous emission rate (see, for example, \cite{Foresi97,NFA07}). This is due to the change of the dispersion relation of the photons and of the photonic density of states due to the environment, that can also yield a photonic bandgap, where the density of states is very small, as well as frequency ranges with peaks of the density of states. Recently, it has also been shown that photonic crystals are very good candidates to enhance and control radiation-mediated forces between atoms or molecules such as the resonance interaction \cite{IFTPRP14,NPR15} and the dipole-dipole interaction \cite{EJ13,GHCCK15}.

Dynamical (modulated) environments, whose  optical properties are periodically modulated in time, have been recently obtained. Oscillating boundaries have been usually considered in the framework of the dynamical Casimir \cite{Dodonov10} and Casimir-Polder \cite{VP08,MVP10,ABCNPRRS14} effects, as well as modifications of these interactions in the case of a fixed boundary \cite{PS07}, or boundaries with quantum fluctuations of their position \cite{BP13,AP15}. The modification of the spontaneous emission process of an atom near an oscillating plate has been also recently investigated using a simple model to quantize the electromagnetic field in the presence of the moving boundary \cite{GHDBZ10}.
The modulation setup we consider in this paper is completely different from that in \cite{GHDBZ10} (oscillating conducting plane boundary), due to the presence of a photonic bandgap and a quadratic dispersion relation near the gap edge. The presence of a modulated band gap, whose edge oscillates in time, is an important feature of our system yielding additional features to the emission spectrum.
A general dynamical control of the decay of a state coupled with a continuum has been considered in Ref. \cite{KofmanKurizki01}.
Important examples of dynamical environments are modulated photonic crystals \cite{Notomi10} and optomechanical crystals \cite{painter} whose optical properties are periodically modulated in time.

New effects arise in the modulated case, for example amplification of light \cite{Ueta12}. Ultrafast control of spontaneous emission in a cavity, on time scales faster than the spontaneous lifetime, has been also achieved \cite{JJSHMVF14}, as well as the control of spontaneous emission of a quantum dot by tuning a photonic crystal cavity \cite{Midolo12}. A modulation frequency of photonic crystal nanocavities in the range of gigahertz has been obtained using acoustic phonons \cite{Fuhrmann11}.

In this paper, we show that new possibilities for controlling and tailoring spontaneous emission exist when the structured environment becomes dynamics, i.e. when its properties are modulated in time through a periodic external action. A relevant case we consider in detail is that of a dynamical photonic crystal. After introducing a model for a dynamical photonic crystal, we show that a dynamical bandgap exists, and consider the effect on the spontaneous emission of a two-level quantum emitter.
More specifically, we show that not only a modification of the emission rate is obtained, but that also the emission spectrum can be modified and controlled through the modulated environment. We show that two side peaks appear in the emission spectrum of the atom, whose distance from the central peak is related to the modulation frequency of the environment.
We discuss how this feature, specific of our model of modulated photonic crystal, could be exploited to obtain the density of states of the environment from the emission spectrum of a quantum emitter of fixed frequency by sweeping the environment's modulation frequency.
The two peaks are not symmetric in height, due to the peculiar photon density of states of the photonic crystal, which is different at the frequencies of the two peaks.
This clearly shows how a dynamical environment, even in the adiabatic regime we consider, can allow one to obtain a qualitative change of the emission spectrum and gives new possibilities to tailor the features of the spontaneous emission process of the quantum emitter.
We point out that the dynamical environment we consider has physical features very different from the oscillating reflecting plate considered in \cite{GHDBZ10}, being characterized by a completely different dynamical dispersion relation yielding a dynamical photonic-band-gap and dynamical quadratic dispersion relation.

This paper is organized as follows. In Sec. \ref{sec:DynamicalEnvironment} we introduce our model of a modulated photonic-band-gap environment, obtaining explicitly the dynamical dispersion relation and gap frequencies for a modulated photonic crystal in the adiabatic hypothesis. In Sec. \ref{sec:spontaneous emission} we investigate the emission spectrum of a two-level quantum emitter embedded in our modulated band-gap environment, and discuss its main physical features. Section \ref{conclusions} is dedicated to our conclusive remarks.

\section{The model of time-modulated bandgap environment}
\label{sec:DynamicalEnvironment}

A photonic-band-gap environment possesses a forbidden gap in the photon frequencies, where photons cannot propagate because the density of states vanishes. In the region external to the gap, and in the proximity of its edge, the dispersion relation in the  effective mass approximation has the following quadratic form \cite{LMS15,JW90,JW91}
\begin{equation}
\wk = \omega_g \pm A(k-k_0)^2 ,
\label{staticdispersionrelation}
\end{equation}
where $\omega_g = \omega_l \, , \omega_u$ is the frequency at the band edge ($\omega_l$ being the lower edge of the gap and $\omega_u$ its upper edge), $k_0$ the wavenumber at which the gap occurs, $A$ is a positive constant, whose value depends on the specific environment considered, and the sign plus or minus refers to frequencies above the upper edge or below the lower edge, respectively.

In the dynamical case, one can assume that both the gap frequency and the curvature of the parabolic dispersion relation \eqref{staticdispersionrelation} are modulated in time with a frequency $\omega_c$:
\begin{eqnarray}
\omega_g (t) &=& \omega_g + \bar{\xi} \sin (\omega_c t) ,
\label{dyngap} \\
\omega_k(t) &=& \omega_g(t) +A(k-k_0)^2 \left[ 1 + \xi ' \sin (\omega_c t) \right] ,
\label{dyn}
\end{eqnarray}
where $k_0$, $\wg$, $\bar{\xi}$, $\xi '$ and $A$ are constants characterizing the modulated environment. This is the generic kind of dynamical environment we will assume in this paper. We will now develop a model of a one-dimensional modulated photonic crystal and an isotropic three-dimensional one,
yielding a band-edge frequency and dispersion relation in the form given by Eqs. \eqref{dyngap} and \eqref{dyn} with $\xi '=0$; thus, in this specific case the only effect of the modulation is a periodic change of the gap-edge frequency (at the end of this section we will mention other modulated environments where also the curvature of the dispersion relation changes in time).
From the dispersion relation \eqref{dyn}, in the isotropic three-dimensional case we can obtain the following dynamical density of states valid above the gap in the effective mass approximation
\begin{equation}
\rho (\wk (t),t) = \frac V{(2\pi )^3}\frac {k_0^2 \Theta \left[ \wk (t) -\wg (t) \right]}{2\sqrt{A}\sqrt{\wk (t) -\wg (t)}} \left( 1- \frac {\xi '}2 \sin (\omega_c t) \right) ,
\label{dyndensst}
\end{equation}
where $\Theta (x)$ is the Heaviside function. Equations \eqref{dyngap}, \eqref{dyn} and \eqref{dyndensst} characterize our dynamical environment. We wish to stress, however, that our results are not limited to such a case, being valid for any periodically modulated photonic-bandgap-environment with a quadratic dispersion relation and a periodically driven edge frequency (a coupled cavities array with a time-dependent hopping constant, for example).

\begin{figure}
\includegraphics[width=0.48\textwidth]{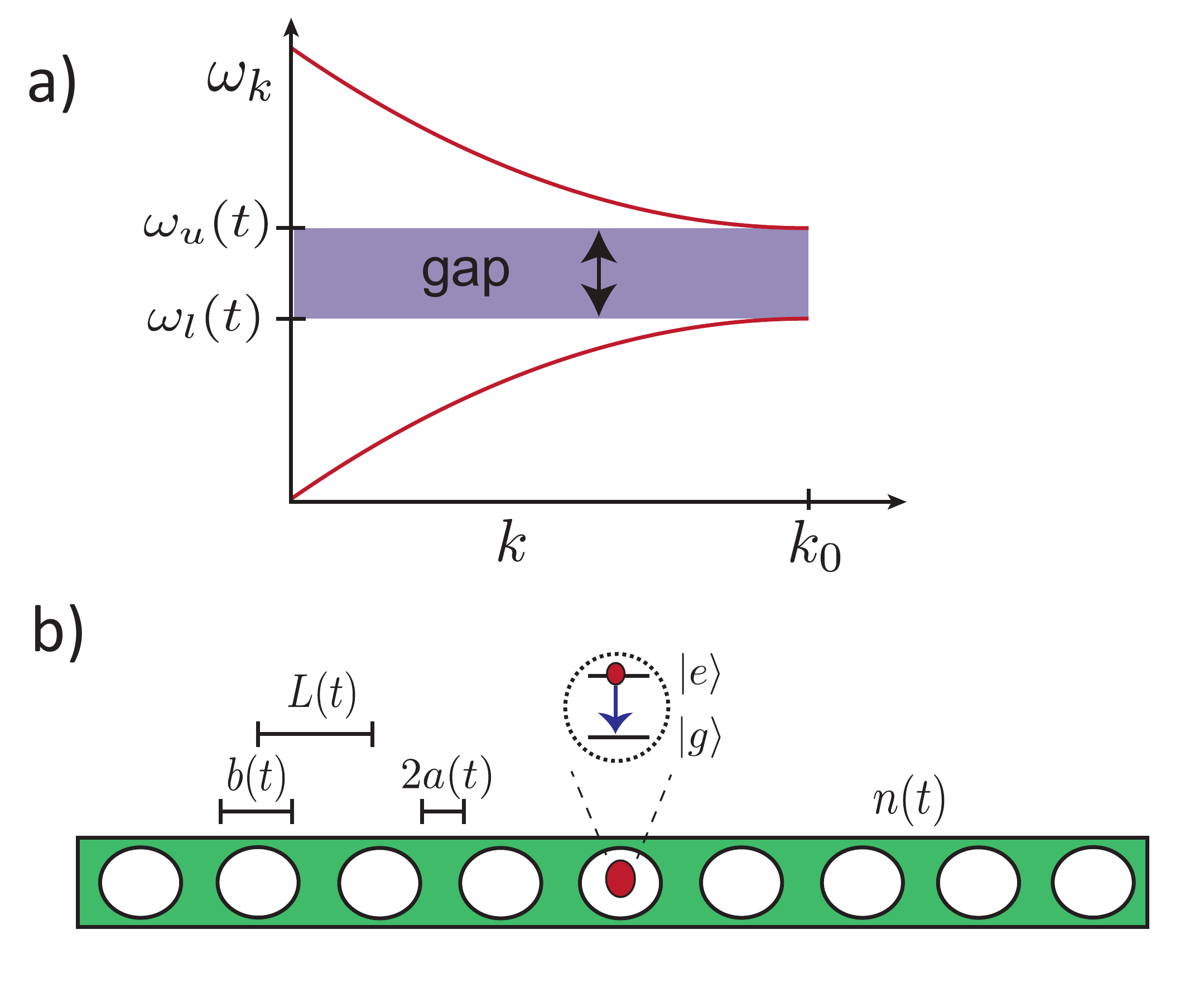}%
\caption{(a) Typical dispersion relation in a photonic band gap system for frequencies close to the band edge. The dispersion relation for $k>k_0$ has been translated to the first Brillouin zone. In our set up the modulation of the environment induces oscillations of the lower and upper edge frequencies. (b) Sketch of a time-modulated bandgap system such as a photonic crystal, where the refractive index or the lattice constants depend on time.}
\label{set_up}
\end{figure}

Our model of a modulated photonic crystal consists of a periodic sequence of dielectric slabs whose refractive index or lattice constant depend periodically on time, within the effective mass approximation.
It is an extension of the static model introduced in Refs. \cite{JW90,JW91} and it is illustrated in Fig.\ref{set_up}. It consists of dielectric slabs of thickness $2a$ made of a nondispersive and non dissipative dielectric with refractive index $n$, separated by a distance $b$ of vacuum space; $L=2a+b$ is the periodicity of the crystal in one direction, while it is homogeneous in the two other directions. In the static case, the implicit dispersion relation is \cite{AKP04}
\begin{eqnarray}
\cos (kL) &=& \cos \left( \frac {\wk b}c \right) \cos \left( \frac {2\wk na}c \right) \nonumber \\
&-& \frac {n^2 +1}{2n} \sin \left( \frac {\wk b}c \right) \sin \left( \frac {2\wk na}c \right) .
\label{impldisprelstatic}
\end{eqnarray}
This relation can be inverted by assuming $b=2na$, obtaining \cite{AKP04}
\begin{equation}
\wk = \frac c{4na} \arccos \left[ \frac {4n \cos [2ka (1+n)]+(1-n)^2}{(1+n)^2} \right] .
\label{disprelstatic}
\end{equation}
This relation gives frequency gaps at $k=m\pi /L$, with $m$ integer, and Fig.\ref{set_up}(a) shows the dispersion relation and the first gap at $k_0= \pi /L$. In the proximity of the band edges, the effective mass approximation can be used, yielding the dispersion relation \eqref{staticdispersionrelation}, where the frequency gap edges $\wg = \omega_\ell , \omega_u$ and the constant $A$ can be expressed in terms of the physical parameters of the crystal \cite{JW90,JW91,AKP04,LNNB00}, even if it can overstimate the density of states in the proximity of the band edges \cite{LNNB00}.
In the isotropic three-dimensional model, this dispersion relation is assumed valid regardless of the direction of propagation of the photon \cite{JW90}. This isotropic model has been frequently used in the literature and proved to be a good model of a real photonic crystal for studying many radiative processes in a structured environment \cite{EJ13,JQ94,WGWX03}, even if it can overestimate the density of states in the proximity of the band edges \cite{LNNB00}.

In our dynamical (modulated) extension, we assume that some generic system parameter $r$, for example the refractive index of the dielectric slabs or the lattice constant (see Fig. \ref{set_up}), depends periodically on time according to the following law,
$r(t)= r_0 \left[ 1 + \xi \sin (\wc t) \right]$,
where $\wc$ is the oscillation frequency of the parameter, $r_0$ its average value and $\xi \ll 1$ the oscillation amplitude, assumed small.
Such  modulation can be experimentally obtained, for example, by exploiting acoustic phonons in the medium, as in Ref. \cite{Fuhrmann11}, or by inducing mechanical vibrations in the structure \cite{painter}.

We first assume that the modulation of the crystal is obtained by a periodic modulation of the refractive index of the dielectric slabs, and we derive the dynamical dispersion relation, the frequency gaps and the density of states. Similar results are also obtained in the case of a periodic modulation of the lattice constants ($a$ and $b$), as we will discuss later on. The time-dependent refractive index is
\begin{equation}
n(t) = n_0[1+\xi (t)]=n_0 [1+ \xi \sin (\wc t) ]
\label{modrefractindex}
\end{equation}
with $\xi (t) = \xi \sin (\wc t)$, and where $n_0$ is its average value and $\wc$ is the modulation frequency.

The Maxwell wave equation for the electric field is
\begin{equation}
\nabla^2 \bE = \frac 1{c^2} \frac {\partial^2}{\partial t^2} \left[ \epsilon (x,t) \bE \right] \, ,
\label{waveeq}
\end{equation}
where $\epsilon (x,t)$ is the time-dependent dielectric constant of the slabs of the photonic crystal, which depends on $x$ according to the periodicity of the photonic crystal and on time due to the periodic modulation of the refractive index $n(t)$. Terms containing time derivatives of $\epsilon (x,t)$ appear in \eqref{waveeq}, eventually leading to the dynamical Casimir effect \cite{Dodonov10} or amplification of light \cite{Ueta12}. It is not easy to find solutions of \eqref{waveeq} with the boundary conditions given by the photonic crystal \cite{Ueta13}; however, assuming a slowly varying modulation of the dielectric constant, we can neglect the time derivatives of $\epsilon (x,t)$ in \eqref{waveeq}. This adiabatic approximation consists in assuming that $\wc$ is much smaller than any other characteristic frequency involved in the problem, in particular the atomic transition frequency and field frequencies relevant for the process considered. Real photons emission by the dynamical Casimir effect is thus neglected.

Under the hypothesis of the adiabatic modulation mentioned above, the implicit dispersion relation is the same as that of the static one given in \eqref{impldisprelstatic} with the substitution of $n$ with the time-dependent refractive index $n(t)$.
Thus our implicit dispersion relation becomes
\begin{eqnarray}
\cos (kL) &=& \cos \left( \frac {\wk b}c \right) \cos \left( \frac {2\wk a n_0[1+ \xi (t)]}c \right) \nonumber \\
&-& \frac {n_0^2[1+\xi (t)]^2 +1}{2n_0(1+\xi (t))} \sin \left( \frac {\wk b}c \right) \nonumber \\
&\times& \! \! \sin \left( \frac {2\wk an_0[1+\xi (t)]}c \right) .
\label{impldisprelmod}
\end{eqnarray}

The usual assumption $2na=b$, used to invert Eq. \eqref{impldisprelstatic} in order to obtain an explicit dispersion relation as in \eqref{disprelstatic}, cannot be used in our modulated case because $n(t)$ depends on time, while $a$ and $b$ are fixed. We thus expand \eqref{impldisprelmod} at first order in $\xi$, assuming a small perturbation of the photonic crystal [$\xi (t), \xi \ll 1$] and $2n_0a=b$. After straightforward algebraic calculations, we obtain

\begin{eqnarray}
\label{moddisprel}
\cos (kL) &\simeq& \cos^2 x -\frac {n_0^2+1}{2n_0} \sin^2 x -\left[ \left( 1+ \frac {n_0^2+1}{2n_0}  \right)  \right. \nonumber \\
&\times&\! \! \left. x \sin x \cos x +\frac {n_0^2-1}{2n_0} \sin^2 x \right] \xi (t) ,
\end{eqnarray}
where $x = 2\wk (t) n_0a/c$ is a dimensionless quantity that in general depends on $t$ through $\wk (t)$. When \eqref{moddisprel} is inverted to obtain the explicit dispersion relation, it shows frequency gaps due to the multivalue character of the $\arccos (y)$ function, analogously to the well-known static case. We can now expand \eqref{moddisprel} at first order in $\xi (t)$ around the value of $x(t)$ corresponding to the static case $x_g = 2\wg n_0a/c$, $\wg =(c/4an_0) \arccos [(1+n_0^2-6n_0)/(1+n_0^2+2n_0)]$ being the lower edge frequency obtained in the static case. For the first gap at $k_0=\pi /L$, we obtain the following expression of the frequency of the edge of the dynamical gap
\begin{equation}
\wg (t) = \wg + \frac c {2n_0a} \frac \alpha \beta \xi \sin ( \wc t),
\label{dyngapfreq}
\end{equation}
where
\begin{eqnarray}
\alpha &=& - \frac {(n_0+1)^2}{2n_0} x_g \sin (2x_g) - \frac {n_0^2-1}{2n_0} \sin^2 x_g , \nonumber \\
\beta &=& \frac {(n_0+1)^2}{2n_0}\sin (2x_g) .
\label{alfabeta}
\end{eqnarray}
Equation \eqref{dyngapfreq} is in the form \eqref{dyngap} with $\bar{\xi}= c\alpha /(2n_0a\beta )$.
Also, expanding the left-hand side of \eqref{moddisprel} around $k_0=\pi /L$, we can obtain the dynamical dispersion relation in the proximity of the band edge (effective mass approximation),
\begin{equation}
\wk (t) = \wg (t) + A (k-k_0)^2 ,
\label{dyndisprel}
\end{equation}
where $\wg (t)$ is given by \eqref{dyngapfreq} and the coefficient $A$ by
\begin{equation}
A = \frac {cL^2}{4n_0a\beta}=  \frac {cL^2}{2a} \frac 1{(n_0+1)^2 \sin \left( \frac {4n_0a\wg}c \right)} .
\label{coeffA}
\end{equation}
The results \eqref{dyngapfreq} and \eqref{dyndisprel} yield a dynamic gap frequency and dispersion relation in the form \eqref{dyngap} and \eqref{dyn}, respectively, with $\xi ' = 0$. This shows that, at the order considered in $\xi$, the effect of the modulation of the photonic crystal is a periodic change of the gap frequencies. Similarly to the usual static case, the isotropic model we will use in the following extends this dispersion relation to the three-dimensional case, assuming it valid independently of the propagation direction of the photon. In the next section we will investigate the effect of the modulation of the photonic crystal on the spontaneous emission of a quantum emitter such as an initially excited atom or quantum dots, in particular on the emission spectrum.

Similar results can be obtained if the modulation involves one or more lattice constants. For example,
let us assume a time modulation of the lattice constant $L(t)=b(t)+2a(t)=L_0[1+\eta \sin (\omega_c t)]$ (see Fig. \ref{set_up}), keeping constant the refractive index $n$. Under the hypothesis of an adiabatic modulation, the implicit dispersion relation reads
\begin{eqnarray}
\cos [kL(t)]&=&\cos\frac{\wk b(t)}{c}\cos \left( \frac{2\omega na(t)}c \right) - \frac{n^2+1}{2n}\nonumber \\
&\times&  \! \! \sin \left( \frac{\wk b(t)}c\right) \sin \left( \frac{2\wk na(t)}c\right).
\label{impldisprelation2}
\end{eqnarray}

Contrarily to the previous case of a time-dependent refractive index, in this case it is possible to take the assumption $2na(t)=b(t)$ by an appropriate choice of a simultaneous modulation of the constants $a$ and $b$, allowing one to simplify the dispersion relation as in the static case. Thus,  in the effective mass approximation we immediately get
\begin{equation}
\wk (t)=\wg (t)+A(t)(k-k_0)^2 ,
\label{dyndisprel2}
\end{equation}
where, working at the first order in $\eta$,
$\wg (t)=\wg [1-\eta \sin (\omega_c t)]$ and
$A(t)=A[1+\bar{\eta}\sin (\wc t)]$ with $\bar{\eta}=\eta [1-\frac{4\omega n a}{c}\cot\left(\frac{4\omega n a}{c}\right) ]$. We thus again recover a dispersion relation of the form \eqref{dyn}.
The isotropic dynamical three-dimensional model we will consider in the next section is the extension of the one-dimensional model introduced in this section, assuming that the dispersion relation is valid for any propagation direction of the photon. The static isotropic model has proven to be a reliable model of realistic three-dimensional photonic crystals, in particular for processes involving frequencies not too close to the edge of the band-gap, as those we are considering \cite{LNNB00} (at frequencies very close to the bandgap edge it overestimates the contribution of these field modes, due to the Van Hove divergence in the density of states).

To conclude this section, we would like to mention that a similar dispersion relation can be also obtained if a tight-binding model of coupled cavities is considered. In this case the time-dependent parameter is the hopping constant $J(t)$ that leads to the dispersion relation $\omega_k(t)=\omega_{cav}-J(t)\cos(k a)$ where $\omega_{cav}$ is the cavity frequency and $a$ the lattice constant. If only one of the two edges is considered, it is possible to perform an effective mass approximation and the dispersion relation reduces again to the form given in \eqref{dyngap} and \eqref{dyn}.

\section{Spontaneous emission in the modulated photonic bandgap environment}
\label{sec:spontaneous emission}

We now consider the spontaneous emission of a two-level atom (or any other quantum emitter) embedded in the dynamical bandgap environment described in the previous section, with special reference to the dynamical photonic crystal; $\mid e \rangle$ and
$\mid g \rangle$ are respectively its excited- and ground-state energy levels separated by an energy $\hbar \wo$. The Hamiltonian, in the minimal coupling scheme and in the dipole approximation \cite{CPP95}, and within our adiabatic approximation, is
\begin{eqnarray}
H&=&H_{atom} +\sum_{\bk \lambda} \hbar \wk (t) a_{\bk \lambda}^\dagger a_{\bk \lambda}
-\frac em \sum_{\bk \lambda} \left( \frac{2\pi \hbar}{\wk (t) V} \right)^{1/2} \nonumber \\
&\times& \! \!
(\ekl \cdot \bp ) \left( a_{\bk \lambda} e^{-i\wk (t)t} + a_{\bk \lambda}^\dagger e^{i\wk (t)t} \right) ,
\label{Hamiltonian}
\end{eqnarray}
where $H_{atom}$ is the Hamiltonian of the atom located at $\bR=0$, $\bp$ the atomic momentum operator and $\lambda$ is the polarization index. The mode frequencies $\wk (t)$ explicitly depend on time due to the presence of the external modulated environment. We assume the atomic transition frequency $\wo$ above the upper edge of the (average) photonic gap $\wg = \omega_u$ and enough far from the edge so that a weak-coupling approach can be applied [in other words, sufficiently far from the divergence of the density of states \eqref{dyndensst} at the band edge]. Thus, the divergence of the density of states at the band edge, typical of the one-dimensional or the isotropic three-dimensional photonic crystal, does not play a significant role in our case; this is obtained when $\wo - \wg \gg 2\wo^{7/2} \! \mid \bd_{eg} \mid^2 \! /(3 \hbar c^3)$,
$\bd_{eg}=-ie/(m\wo )\bp_{eg}$ being the matrix element of the atomic dipole moment operator $\bd$ between the excited and ground states \cite{JQ94}.

An analogous Hamiltonian holds in a one-dimensional case, for example when the quantum emitter is placed in a photonic waveguide \cite{JMW95}, and relevant field modes have a frequency above the cutoff frequency of the guide. Recent experiments have shown striking possibilities of engineering radiative processes of atoms or quantum dots exploiting a photonic crystal waveguide; for example, superradiant emission \cite{Goban15} and atom-atom interactions \cite{Hood16}. In the one-dimensional case the wave vector $\bk$ is along the axis of the guide.

We now consider our modulated (dynamical) case, specifically that described by the dispersion relation \eqref{dyndisprel}, extended to the isotropic three-dimensional case. As mentioned, in this case and at first order, the modulation of the environment has one main effect, that is, a time dependence of the gap frequency, while the curvature of the dispersion relation in the effective mass approximation is constant and equal to the static case. We will show that this yields striking consequences on the spontaneous decay of an atom embedded in the modulated environment, in particular on the spectrum of the radiation emitted. This indicates a further possibility to control the spontaneous emission process through the modulation of the environment, in comparison to a static one.

In our adiabatic hypothesis (slow oscillation frequency of the environment with respect to the atomic transition frequency and relevant field frequencies), the solution of the Heisenberg equation  for the field annihilation operator is
\begin{equation}
\akl (t) = \akl (0) e^{-i[\wk t + \phi (t)]} ,
\label{annsol}
\end{equation}
where
\begin{equation}
\phi (t) = \int_0^t dt' \bar{\xi} \sin (\wc t')
\label{phase}
\end{equation}
is a phase factor arising from the adiabatic time-dependence of the dispersion relation. Assuming the atom in its excited state $\mid e \rangle$ at $t=0$ and using first-order time-dependent perturbation theory, the transition probability from the excited to the ground state of the atom, using Eq. \eqref{annsol} and the dynamical dispersion relation given above, after some algebraic calculations, is obtained as
\begin{eqnarray}
P(t) &=&  \frac {2\pi \wo^2}{\hbar V} \sum_{\bk \lambda} \mid \ekl \cdot \bd_{eg} \mid^2 \nonumber \\
&\times& \left| \int_0^t \! dt' \frac 1{\sqrt{\wk (t')}} e^{i(\wk -\wo )t' } e^{i\phi (t')} \right|^2 .
\label{probability}
\end{eqnarray}

When compared to the analogous quantity in the static case, the main differences in \eqref{probability} are the time dependence of $\wk(t')$ and the presence of the phase factor $\phi (t')$ inside the time integral. Assuming a small amplitude of the modulation of the environment, we can expand the square root of $\wk (t')$ and the exponential function with the phase factor $\phi (t')$ at the first order in $\bar{\xi}$. This allows us to perform analytically the time integral in \eqref{probability}.
Taking into account that $\wc \ll \wk, \wg$ and keeping only terms up to the first order in $\bar{\xi}$,
we can perform the integral in \eqref{probability}, and we get
\begin{widetext}
\begin{eqnarray}
P(t) &=& \frac {2\pi e^2}{\hbar m^2 V} \sum_{\bk \lambda} \mid \ekl \cdot \bp_{eg}\mid^2
\frac 1\wk \left| \frac {2\sin \left( \frac {(\wk -\wo)t}2 \right)}{\wk -\wo}
\right. \nonumber \\
&+& \left. \frac {i\bar{\xi}}{\wc} \left[ - \frac {2\sin \left( \frac {(\wk -\wo )t}2\right)}{\wk -\wo}+
e^{i\wc t/2} \frac {\sin \left( \frac {(\wk -\wo +\wc )t}2 \right)}{\wk -\wo +\wc}
+ e^{-i\wc t/2}
\frac {\sin \left( \frac {(\wk -\wo -\wc )t}2 \right)}{\wk -\wo -\wc} \right]
\right|^2 .
\label{prob1}
\end{eqnarray}
\end{widetext}

In the isotropic three-dimensional case and in the continuum limit $V \to \infty$, after polarization sum, angular integration, introducing the appropriate density of states given in our case by \eqref{dyndensst} with $\xi ' =0$, and using

\begin{eqnarray}
&\ & \sum_{\bk \lambda} \rightarrow \int d\wk (t) \rho (\wk (t),t) \sum_\lambda \int d\Omega ,\nonumber \\
&\ & \sum_\lambda \int d\Omega \mid \ekl \cdot \bp_{eg}\mid^2 = \frac {8\pi}3 \mid \bp_{eg}\mid^2 ,\nonumber \\
&\ & \rho (\wk (t),t) = \frac V{(2\pi )^3}\frac {k_0^2 \Theta \left[ \wk (t) -\wg (t) \right]}{2\sqrt{A}\sqrt{\wk (t) -\wg (t)}} ,
\label{contlim}
\end{eqnarray}
after some algebra and taking into account that in our case the time-dependence in \eqref{dyndensst} cancels out, we finally obtain
\begin{equation}
P(t) = \int_\wg^\infty P(\wk ,t) d\wk ,
\label{probdyn}
\end{equation}
where $P(\wk ,t)$ represents the probability density of emission at time $t$ of a photon with frequency $\wk$, given by

\begin{widetext}
\begin{eqnarray}
P(\wk ,t) &\simeq& \frac {k_0^2 e^2 \mid \bp_{eg}\mid^2}{3\hbar m^2 \sqrt{A}} \frac 1{\sqrt{\wk -\wg}}  \frac 1\wk
\left| \frac {2\sin \left( \frac {(\wk -\wo)t}2 \right)}{\wk -\wo}
\right. \nonumber \\
&+& \left. \frac {i\bar{\xi}}{\wc} \left[ - \frac {2\sin \left( \frac {(\wk -\wo )t}2\right)}{\wk -\wo}+
e^{i\wc t/2}\frac {\sin \left( \frac {(\wk -\wo +\wc )t}2 \right)}{\wk -\wo +\wc}
+ e^{-i\wc t/2}
\frac {\sin \left( \frac {(\wk -\wo -\wc )t}2 \right)}{\wk -\wo -\wc} \right]
\right|^2 .
\label{probdensity}
\end{eqnarray}
\end{widetext}

The first term in the squared modulus in Eq. \eqref{probdensity} does not depend on the environment modulation, and its explicit evaluation yields, after integration over $\wk$ and for large times ($t \gg \wo^{-1}$), the same decay probability, proportional to $t$, of the static case \cite{JQ94}
\begin{equation}
P^{stat} (t) \simeq \frac{2k_0^2 e^2 \mid \bp_{eg} \mid^2 t}{3\hbar m^2 \sqrt{A} \wo \sqrt{\wo - \wg}} .
\label{probabstatic}
\end{equation}

The terms proportional to $\bar{\xi}$ in \eqref{probdensity} give the change to the transition probability due to the modulation of the environment. The long-time limit of expression \eqref{probdensity} gives the spectrum of the radiation emitted by the atom, $S(\wk )=\lim_{t\to\infty}{P(\wk ,t)}$.
Our results are also valid in the case of modulation of the lattice length constants, rather than the refractive index, as well as in the case of a coupled cavities array, when the dynamical dispersion relation is respectively given by \eqref{dyndisprel2} or \eqref{dyn}. In fact, it is possible to show that, within our assumptions, the effect of the time dependence of the constant $A(t)$ is negligible, and thus the effect of the modulation is mainly the periodical change of the gap-edge frequency, as in the case explicitly considered in this section.
An expression similar to \eqref{probdensity} (differing only for constant factors) can be obtained also for the one-dimensional case, so all the results and physical considerations that follow hold in that case too, for example, for an atom or quantum dot in a photonic crystal waveguide, provided its transition frequency is above the cutoff frequency of the guide.

\begin{figure}
\includegraphics[width=0.48\textwidth]{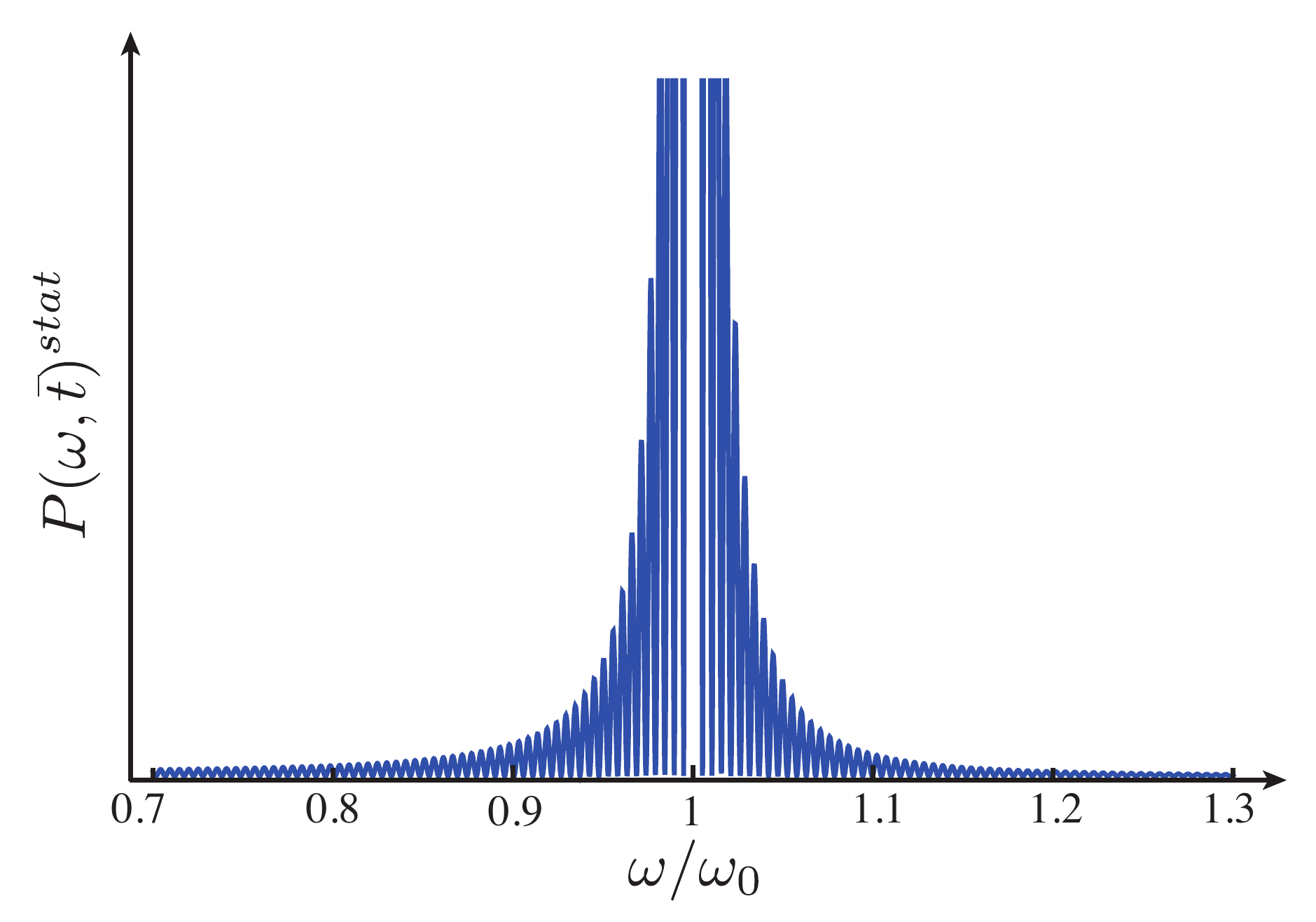}
\caption{Emission probability for unit frequency (arbitrary units) of a photon with frequency $\omega$ at time $\bar t=1200 \wo^{-1}$ in the static case. We have used $\wg /\wo = 0.5$.}
\label{Figure2}
\end{figure}

\begin{figure}
\includegraphics[width=0.48\textwidth]{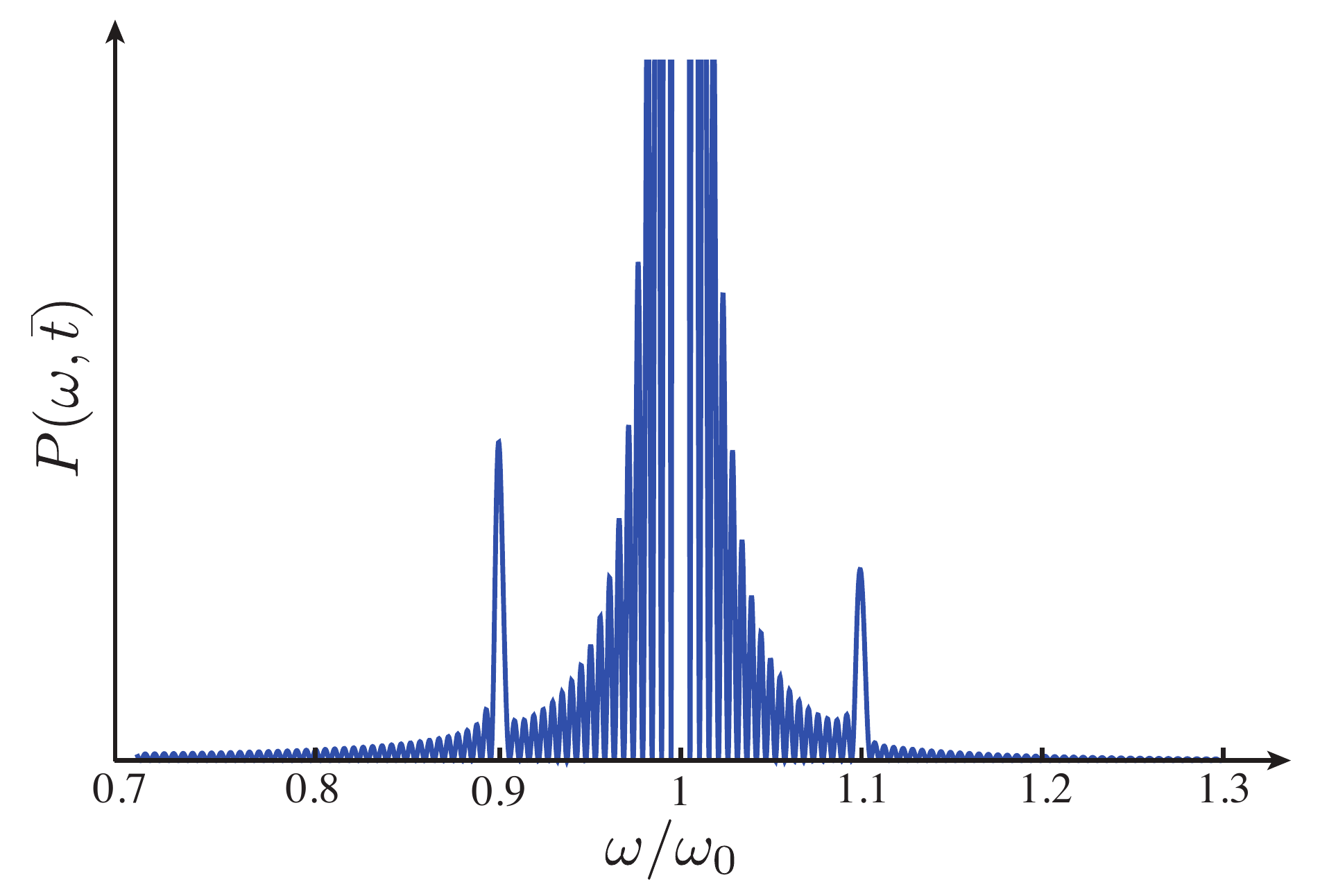}
\caption{Emission probability for unit frequency (arbitrary units) of a photon with frequency $\omega$ at time $\bar t=1200 \wo^{-1}$ in the dynamical case. We have used the following parameters: $\wc / \wo = 0.1$, $\wg /\wo = 0.5$ and $\bar{\xi} =0.01 \wo$.  The figure shows clearly two side peaks at frequencies $0.9 \wo$ and $1.1 \wo$, that is $\omega \simeq \wo \pm \wc$. The asymmetry between the two lateral peaks is due to the different density of states of the bandgap environment at frequencies $\omega \sim \wo \pm \wc$.}
\label{Figure3}
\end{figure}
Figures \ref{Figure2} and \ref{Figure3} show the emission probability for unit frequency obtained from \eqref{probdensity}, respectively for the static ($\bar{\xi}=0$)
and dynamical cases, obtained from (\ref{probability}) after polarization sum and angular integration, at time $\bar t = 1200 \wo^{-1}$. In the dynamical case, we clearly observe, apart from the usual peak at $\omega \sim \wo$, the presence of two side peaks in the radiation emitted at frequencies $\omega \sim \wo \pm \wc$
related to the presence of the energy denominators containing $\omega -\wo \pm\wc$ in \eqref{probdensity}.
It should be noted that the widths of the central and side peaks result from having plotted the emission probability at a finite time, and are not related to the natural width of the spectral lines due to our perturbative approach. In the limit $t \to \infty$, the three peaks become more ad more sharply peaked at frequencies $\wo- \wc , \wo , \wo+\wc$.

The presence of the two peaks in the emission spectrum is our main result on the features of the spontaneous emission in the dynamical environment. They are shifted from the atomic frequency $\wo$ by the modulation frequency $\wc$ of the environment. They are asymmetric, due to the different density of states at their frequency. This effect is a sort of beat between the two natural frequencies involved in the system, $\wo$ and $\wc$,
as one could expect from general considerations based on the Floquet theory of systems periodic in time. The two peaks, which are reminiscent of the Stokes and anti-Stokes lines in quantum optomechanics \cite{AKM14}, are not symmetric in height, due to the different density of states of the photonic crystal at the frequencies of the two peaks; this asymmetry can be enhanced by making the atomic frequency closer to the edge of the bandgap. This is a peculiar feature of our dynamical photonic crystal environment, not present in other cases, for example an oscillating reflecting plate.
The ratio between the heights of the two peaks at $\wk = \wo \pm \wc$ can be obtained from \eqref{probdensity}, and is given by
\begin{equation}
\frac {P(\wk=\wo -\wc )}{P(\wk=\wo +\wc )} \simeq \frac {\rho (\wo -\wc )}{\rho (\wo +\wc )}
= \sqrt{\frac{\wo +\wc -\wg}{\wo -\wc -\wg}} ,
\label{peaks}
\end{equation}
with $\wc \ll \wo ,\wg$, and where $\rho (\omega )$ is the static density of states (or the average dynamical one)
\begin{equation}
\rho (\omega ) = \frac V{(2\pi )^3}\frac {k_0^2 \Theta \left( \omega -\wg \right)}{2\sqrt{A}\sqrt{\omega -\wg}} .
\label{static density states}
\end{equation}
This relation linking the height of the lateral peaks to the density of states at their frequency shows the possibility of investigating the density of states of the environment by sweeping the modulation frequency. In Fig. \ref{Figure3}, the left peak is higher than the right one because it is at a frequency closer to the edge of the band gap, where the density of states is higher. This is a peculiar feature occurring for a modulated environment with a photonic bandgap. Thus, from our results we expect that we can experimentally explore the density of states of a generic bandgap environment, and from that its dispersion relation, by measuring the emitted spectrum of a quantum emitter of fixed frequency inside it, by sweeping the modulation frequency of the environment under investigation. This important feature is not present in the case of an atom in front of an oscillating mirror discussed in \cite{GHDBZ10}, and it is a distinctive point of the modulated environment we have considered.

In order to observe the two lateral peaks they must be separated in frequency from the central peak by more than the natural width of the excited level. Assuming a typical natural width of $\sim 10^{8}$ Hz, this means that a modulation frequency $\wc$ of the order of $\sim 10^{9}$ Hz is sufficient to observe this phenomenon. Such a modulation frequency, being much smaller than a typical optical frequency and relevant field frequencies, is fully consistent with the adiabatic approximation we have used. Also, such frequency is in reach of the actual experimental techniques of dynamical photonic crystals \cite{Fuhrmann11,Fan07} or dynamical mirrors \cite{Agnesi09}. We also mention that an explicit integration of \eqref{probdyn} over $\omega$ allows one to obtain the total emission probability.
We can conclude that our results clearly show a noteworthy aspect of the environment's modulation on the spontaneous emission process, specifically that a dynamical environment allows one to control, apart from the decay rate, also the photon spectrum.
For example, it could be exploited to tune processes involving exchange of resonant photons between atoms or molecules, such as the resonant energy transfer between molecules or chromophores \cite{Scholes03,SKKM14}, or the dipole-dipole interaction \cite{EJ13}, as well as to activate or inhibit this kind of process by exploiting a modulation of the environment with an appropriate frequency.

\section{Conclusions}
\label{conclusions}

In this paper we have considered a quantum two-level emitter in a generic modulated photonic bandgap environment. We have developed a model of such an environment, specifically a dynamical photonic crystal, where the refractive index or the lattice constants are periodically modulated; we have obtained the dynamical frequency gap and dispersion relation. We have then shown that the periodic modulation of the environment yields important effects on the spontaneous emission process, compared to the common case of a static environment. Specifically, we have found observable modifications of the emission spectrum, with two side peaks in the photon spectrum. They are separated from the atomic transition frequency by the environment modulation frequency. The two side peaks are not symmetric, being their height different due to the different density of states in the photonic crystal at their frequency.
This is an important feature of our model, not present in the case of the oscillating plane boundary discussed in  Ref. \cite{GHDBZ10}. Also, we show that it could be a way to experimentally investigate the density of states of the environment through the features of the spontaneous emission  a quantum emitter embedded in it, when the modulation frequency of the environment is swept.
Assuming typical experimental values of the parameters involved, we have also shown that this new effect can be observed with actual experimental techniques of dynamical photonic crystals or dynamical mirrors. More generally, our results clearly indicate the impressive potentialities of modulated environments to manipulate, tailor, or activate or inhibit resonant radiative processes.

\begin{acknowledgments}
R.P. and L.R. gratefully acknowledge financial support by the Julian Schwinger Foundation. G.C. acknowledges the financial support by the European Project SIQS (600645), the COST Action NQO (MP1403), and the Austrian Science Fund (FWF) through SFB FOQUS F40, DK CoQuS W 1210 and the START grant Y 591-N16.
\end{acknowledgments}

\end{document}